\documentclass{article}
\usepackage[utf8]{inputenc}
\usepackage[numbers]{natbib}
\usepackage{graphicx}
\usepackage{booktabs}
\usepackage{float}
\usepackage{amsmath}
\usepackage{amssymb}
\usepackage{subcaption}
\usepackage{hyperref}
\usepackage{fancyhdr}

\newcommand{\footremember}[2]{%
    \footnote{#2}
    \newcounter{#1}
    \setcounter{#1}{\value{footnote}}%
}
\newcommand{\footrecall}[1]{%
    \footnotemark[\value{#1}]%
} 

\author{%
  Gefen Kohavi\footremember{contrib}{Indicates equal contribution, authors sorted reverse alphabetically.}
  \and Daniel Ho\footrecall{contrib} 
  \and Michael Gussert
  }

\title{Using Generative Models to Simulate Cosmogenic Radiation}
\date{NVIDIA}

\begin{document}

\maketitle

\begin{abstract}
We introduce HAWCgen, a set of deep generative neural network models, which are designed to supplement, or in some cases replace, parts of the simulation pipeline for the High Altitude Water Cherenkov (HAWC) observatory. We show that simple deep generative models replicate sampling of the reconstruction at a near arbitrary speedup compared to the current simulation. Furthermore, we show that generative models can offer a replacement to the detector simulation at a comparable rate and quality to current methods.  This work was done as part of an undergraduate summer intern project at NVIDIA during the month of June, 2018.
\end{abstract}

\section{Introduction}
The methods examined for this study are from a family of models known as Generative Models.  These methods represent a fundamentally different approach to distribution sampling compared that employed by physicists, as no knowledge of the mathematical form of the expert distribution is needed. The only requirement is that the expert sample is ``large enough" to express the distribution. What is and isn't ``large enough" is a question that is the subject of ongoing research, so developing a functional generative model requires some amount of numerical experimental work to be done.

A question that a physicist might ask is this; ``Why are generative models useful? How can a generative model help us learn something about physics?". While it is true that a generative model can be expressed mathematically, the expression is not useful to a physicist. Such an expression would take the form of something akin to a Fourier series, which doesn't necessarily tell you anything useful about what's actually being described by the model. To a physicist, knowing that some distribution is Gaussian or Poissonian is much more useful that being able to draw samples from it. So why use generative models at all?

The key is this; \textit{Monte Carlo based simulations are generative models}. The final output for such a simulation might contain many reconstructed parameters, but only a few of those might be useful for analysis. Those few parameters can be thought of as being part of a very high dimensional joint probability distribution. A Monte Carlo based simulation must be manually tuned in order to make these distributions match observations, but a neural network based generative model is end to end differentiable, so this tuning can be done through something akin to gradient descent / ascent.  These types of neural networks are called Generative Networks.

At first glance this might seem a bit backwards; after all, some expert sample must exist in order for a generative model to be trained.  What use is a neural network generative model if the full Monte Carlo simulation is required in order to create it?  

For HAWC \cite{HAWC}, this simulation pipeline is long, complex, and difficult to tune. It takes roughly 11 days to run a full simulation of HAWC on the CPU cluster at the University of Maryland, from the generation of CORSIKA \cite{CORSIKA} showers all the way to the final reconstruction. This is because HAWC is a high duty cycle, high statistics experiment. In order for a simulation to be useful, HAWC requires enough simulated events to do a statistical analysis. At the moment, ``enough simulated events" is around 8 billion in total (3 billion each for gammas and protons, 1.3 billion-ish for helium, and the rest are spread across the other primaries). 

The 1D GAN model described in this document can be trained in less than an hour on a small sample of the simulation. The generation rate of events from these models is on the order of 100x faster than the simulation, or more (no advanced timing studies were done). The idea is then to create the simulation, tune it to physicality, generate a sample, and finally use that sample to generate the remaining statistics extremely rapidly by using a generative model.  Because of how rapidly these statistics can be generated, it becomes possible to answer questions like ``what does the distribution of the number of hits look like for showers of an energy exactly equal to E?" or ``what is the mathematical relationship between zenith angle and the distribution of gamma-hadron separation parameters?".  Answering these questions in the traditional fashion would require either the generation of new simulation, or using some form of reweighting on the current simulation.

The PixelCNN \cite{PixelCNN} model described below represents the state of the art in image generation.  By thinking of the HAWC observatory as a pixel grid, the PixelCNN model can be used to generate a two channel detector sim that is qualitatively indistinguishable from HAWCSIM.  While this model is much more costly to run, and does not offer significant speedup without an investment in new hardware, it shows that it is absolutely possible to completely model the DAQsim with a neural network.  The PixelCNN results suggest that there might be a Generative Adversarial Network (GAN) \cite{GAN} based model to generate the same results at a much faster rate.

In short, generative neural network models can be thought of as highly responsive, automatically tunable approximations for classical Monte Carlo physics simulations.

\subsection{Generative Adversarial Networks}
A GAN is composed of 3 parts; the generator $G$, the discriminator $D$, and the expert sample $\textbf{X}$. Traditionally, the generator and discriminator are neural networks, while the expert sample is a sample of the distribution you are trying to mimic. The generator network takes in a vector of random noise $z\sim$N(0,1) (which it needs as an entropy source) called the latent space, and produces an output that has the same format as that of the expert sample, $G(z)$. The Discriminator takes in a sample (either from the generator, $G(z)$, or a selection from the expert sample $x$), and returns the probability that the sample is a forgery from the generator. As a result, the discriminator effectively acts as a loss function that learns over time. The model parameters of the generator are tuned to minimize $D(G(z))$ while the parameters of the discriminator are turned to maximize $D(G(z))$ and minimize $D(x)$. The end result is a mathematical arms race between $G$ and $D$ (they are ``adversaries"), that slowly moves the distribution $G(z)$ closer and closer to the expert sample $\textbf{X}$. 

The conditional variant of GAN is very similar to a traditional GAN both in terms of architecture and training method. The major difference between the two is that both training and generated data have some label  \cite{CGAN}. In the case of images, an example of this data might be images of cats or dogs with a label of 0 or 1 respectively. This label is concatenated to the input of both the generator and the discriminator. When the discriminator is deciding between real and fake, it can use this extra information in its decision and the generator will be trained to generate images that are expected from that specific label. These labels are not limited to specific classes; they can represent more qualitative data too. In the example of cat or dog generation, this could represent things such as fur color, angle of image, size of animal, and so on.

The basic GAN model can be improved somewhat by modifying the loss function.  The original loss for a GAN is designed to extremize the log-liklihood of the discriminator. 

\begin{equation}
\begin{split}
    L_\textbf{G} &= \mathbb{E}[\log(D(G(z)))] \\
    L_\textbf{D} &= \mathbb{E}[\log(D(\textbf{X}))]+\mathbb{E}[\log(1-D(G(z)))] 
\end{split}
\end{equation}

That said, this loss function has two major problems. First, the useful range (the range where the domain is neither zero or one) of this function is often small. This means that it's easy for the discriminator to become ``certain'' of it's decision, which results in it outputting either zero or one.  When this happens, the gradients vanish and training stops. The second major issue is that the log-liklihood loss is not experimentally meaningful in this scenario because both networks are learning at the same time (the generator or discriminator having a small / large loss doesn't correlate with the quality of the generator output).

An improvement of the traditional GAN is the Wasserstein GAN (WGAN) \cite{WGAN}. The Wasserstein GAN is exactly the same as a traditional GAN except for its loss function. While the loss function of a traditional GAN model extremizes the log-likelihood, the loss function of a WGAN minimizes an approximation of the Earth Mover distance metric also known as the Wasserstein-1 distance.  A full discussion of the Wasserstein-1 distance is beyond the scope of this note, but intuitively this metric is a measure of the minimum amount of "probability mass" that needs to be moved (and how far it needs to be moved) in order to make one distribution look like another.

The benefit of this distance function over the original log-likelihood metric is that any change in distributions will change the Wasserstein distance. This causes the decision boundaries for the discriminator to become linear, and prevents the gradients from vanishing. The new loss functions for the WGAN are:

\begin{equation}
\begin{split}
    L_\textbf{G} &= \mathbb{E}[D(G(z))] \\
    L_\textbf{D} &= \mathbb{E}[D(x)]-\mathbb{E}[D(G(z))]+\lambda\mathbb{E}[P]~.
\end{split}
\end{equation}

Where $\lambda$ represents a tunable constant specified before training and $P$ represents a penalty on the gradient of the discriminator.  These parameters and modifications to the WGAN are described in more detail in \cite{WGANGP}.  This new loss function has resulted in increased stability and reduction in mode collapse (generator outputs the same values). Gradient clipping or a gradient penalty was added to the loss of the discriminator to further increase stability. The result of these changes resulted in a training process where the loss curves of the generator and discriminator are now more meaningful as both loss curves approach zero.

\subsection{Convolutional Neural Networks}
Some data cannot be easily interpreted by fully connected neural networks, and images are a perfect example of this.  For most data, the ordering of the datum is entirely irrelevant:  A vector of (hits, intensity, zenith angle, azimuth angle) contains the same information as a vector of (zenith angle, hits, azimuth angle, intensity). Not so for images.  The value of an individual pixel in an image tells only part of the story; the location of that pixel relative to others in the image tells the rest.  That is to say, in an image, the \textbf{index} of a datum is as important as the datum itself.  This makes images ill suited for fully connected neural networks.  The information stored in an image is relatively invariant with respect to scale or location (a picture of a cat is independent of where the cat is located in the picture).

Convolutional Neural Networks (CNN) were explicitly created to handle data of this form.  Instead of a neuron being fully connected to to the previous and next layer, the neurons of a CNN each have a receptive field known as a kernel.  Each neuron has only a few weights (say, 9, for a 3x3 receptive field), but the kernel is passed over the whole of the previous layer.  At each location, the inner product between the receptive field and the activations from those location on the previous layer is calculated.  This is then passed through some form of non linear function, which produces the activation for the current layer at that position.  In this way, each neuron in a CNN produces an ``image-like" structure known as a feature map, and this process is effectively a numerical convolution over the previous layer, hence the name. 

This does a number of things but first and foremost is that it allows the network to detect features in a transnationaly invariant way that is independent of scale.  The transnational invariance is a result of the kernel convolution, while the scale independence stems from convolutional layers being stacked in a sequence.  Because each value in a feature map is the result of the values at multiple locations in the previous layer, deeper kernels in the network effectively ``see" a larger portion of the image than shallower kernels.  

\subsection{Pixel Convolutional Neural Networks}
Pixel Convolutional Neural Networks (PixelCNN) \cite{PixelCNN} are a class of powerful generative models that achieve state of the art results on many image dataset benchmarks. A PixelCNN model is naturally able to condition on prior information. Such a conditional model can also be thought of as a function between input latent parameters and resulting image samples. There have been many extensions to the original model such as \cite{fastPixelCNN}, \cite{PixelCNNpp} which improve performance, training speed, and sampling speed.

PixelCNN attempts to estimate the joint probability distribution of pixels in an image: given the pixels, the network computes the probability of the image appearing in the distribution represented by the dataset. Commonly, this is the distribution of natural images captured by cameras, including objects like dogs, bikes, and bottles. 

A model computes $p(\textbf{x})$, where $\textbf{x}$ is the image data. We can factorize this probability as a product of conditional distributions over individual pixels:

\begin{equation}
p(\textbf{x}) = \prod_{i=1}^{n^2}p(x_i | x_1, \dots , x_{i_1})
\label{eq:pixelcnn}
\end{equation}

Where $x_i$ is the $ith$ pixel in an ordering of pixels, and the value of pixel $x_i$ is computed given the values of all previously computed pixels. For image data, we can compute the pixels row by row, and pixel by pixel in each row. 

Additionally, we can simply split the generation of each pixel into multiple probabilities for each subpixel:

\begin{equation}
p(x_i) = p(x_{i,R}|x_{<i})p(x_i,G|x_{<i,x_{i,R}})p(x_i,B|x_{<i,x_{i,R},x_{i,G}})
\label{eq:pixelcnn-cond-eqn}
\end{equation}

PixelCNN is an auto-regressive model, in the sense that output from the model (previous pixels) is fed back into the network to compute the next pixel value. This pixel by pixel generation means that we have to propagate through the network a number of times equal to the number of pixels in the image, making it relatively expensive to generate samples. 

PixelCNN++ \cite{PixelCNNpp} simplifies this computation noting that subpixel values, especially red, green, and blue pixel values in commonly used datasets, tend to be very closely related. They save computation by restricting the relationships between channels of a pixel to be linear. Moreover, while PixelCNN outputs a probability distribution over discrete values for each pixel (for images 0 to 255), PixelCNN++ models the pixel distribution with a mixture of logistic distributions and takes the value deemed most likely by the mixture.

In PixelCNN, the probability of each pixel is modeled by a deep, convolutional network. To generate a new pixel, we need to estimate the conditional probability in Equation \ref{eq:pixelcnn-cond-eqn}. We use a network consisting of so-called ``casual convolutions'', where we mask the convolution operation to only ``see'' the values of previously generated pixels. This gives us the ability to condition on previous values without ``cheating'' by looking at future values ahead of time. 


\section{Experiments}

\subsection{Data Preprocessing and Dataset Generation}
Approximately 75,141 events of gamma ray detector sim (about 300 MB) in total were used for these studies.  Of these, 60,112 events were used for training (about 80\%) while the remainder was used for validation.  These simulations were given to us in the form of XCDF files (https://github.com/jimbraun/XCDF) which we converted to .npy files. For our experiments, we use only sims from gamma ray events, although it should be straightforward to apply the models to any type of event.

\begin{figure}[H]
  \centering
    \includegraphics[scale=0.3]{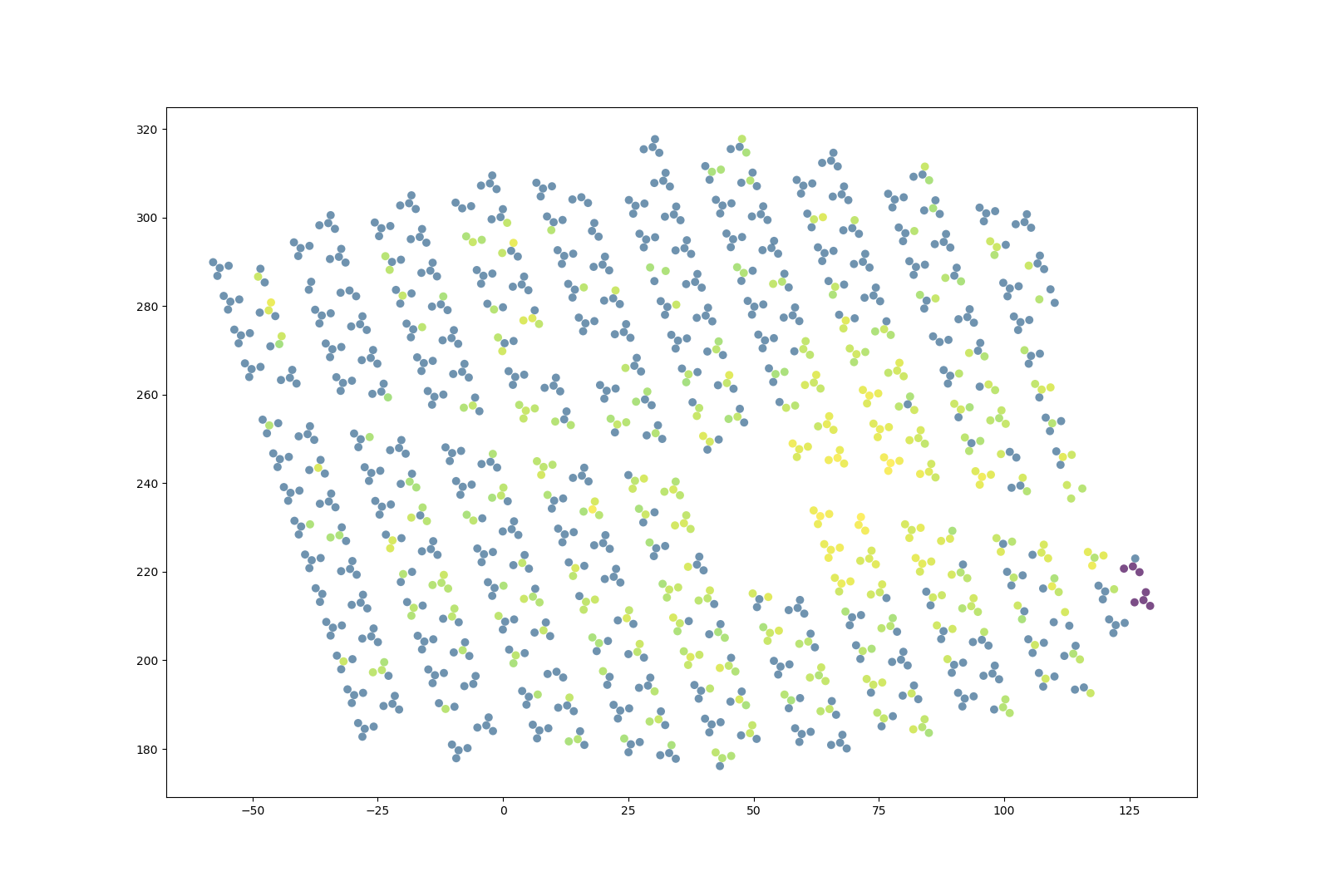}
    \caption{An example of a cosmic ray event in the HAWC grid, from the dataset}
    \label{fig:hawc-grid}
\end{figure}

\begin{figure}[H]
  \centering
    \includegraphics[scale=0.7]{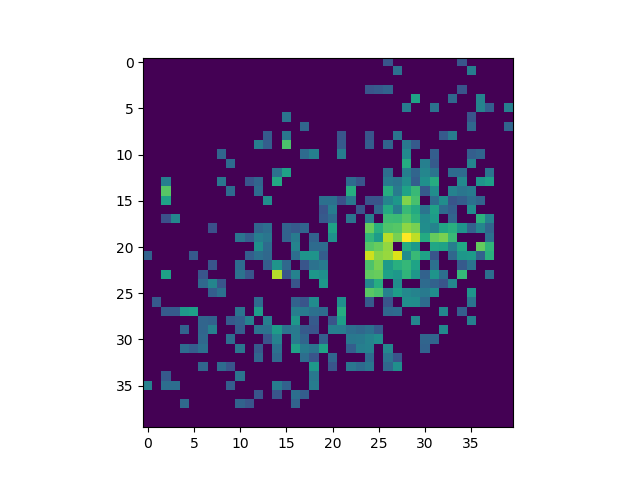}
    \caption{The same event as the one in Figure \ref{fig:hawc-grid} (above), but mapped to a 40x40 array. Color is added for visual clarity.}
    \label{fig:hawc-grid-array}
\end{figure}

In order to work with the raw detector sim, we extracted the charge and hit time values corresponding to each PMT in the observatory, and mapped the data to a $(40, 40, 2)$ array. Each entry in this array corresponds to a PMT location (and either hit or time data). Note that the array will have missing values at certain locations because there are more locations than PMTs. These locations are filled with 0 PEs for the charge channel, and -500 ns for the time channel. Applying this transformation allows us to use 2D convolutional layers in our neural network to exploit local relationships in data of nearby PMTs. Additionally, this method allows us to easily recover the data corresponding to the original PMTs from our array, by applying the reverse transform. This method was proposed by Edna Ruiz in (https://github.com/ednaruiz/ConvoNN).  Finally, we standardized data to the range $[0, 255]$, which allows the array to be interpreted as a standard visual image and lets us start with common, off-the-shelf image models. See \ref{fig:hawc-grid} and \ref{fig:hawc-grid-array} for a visualization of the transform between the grid and the array.

\begin{figure}[H]
  \centering
  \includegraphics[scale=0.3]{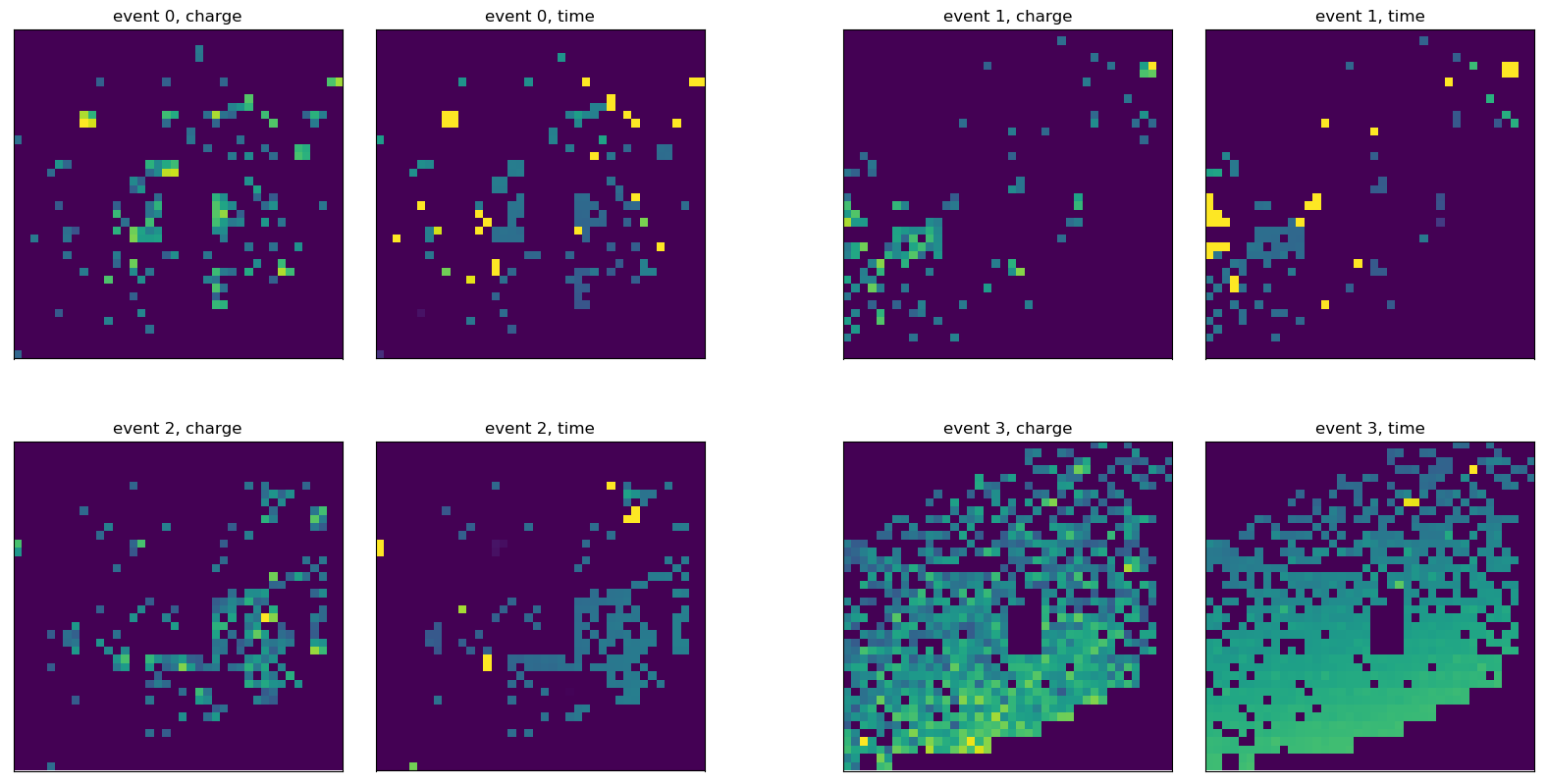}
  \caption{Two Channel "data" from the simulation: channels are displayed with log charge on left and time on right. We can observe some features of the simulation: variations in shower density, center and angle of shower, smooth Gaussian dropoff in charge dimension, smooth linear gradient in time dimension with variations in duration (higher range of color is longer duration), some noise in time dimension with bright yellow spots. Note that some edges of the image and a block in the center does not change in color, as there are 1600 pixels but less than 1200 mapped PMTs. For additional visuals, see the Appendix.}
  \label{fig:ground-truth}
\end{figure}

We visualized each event in Figure \ref{fig:ground-truth}, which is the raw data we fed into models. Note that color has been added for ease of visualization, but each pixel has only 1 dimensional data.

From the original XCDF files, we also pull reconstruction parameters. When used, we scale the data to have a mean of zero and a standard deviation of unity in each dimension.

\subsection{1D Data Generation}
\subsubsection{1D GAN}
Our goal for the 1-dimensional GAN model was to create a efficient way to sample reconstructed simulation parameters. The parameters we chose were  rec.logNPE, rec.nHit, rec.nTankHit, rec.zenith, rec.azimuth, rec.coreX, rec.coreY, and rec.CxPE40 \footnote{these parameters are the log of the number of photoelectrons reconstructed, the number of photomultiplier tubes with signal, the number of tanks with signal, the zenith angle, the azimuth angle, the x locaiton of the reconstructed shower core, the y location of the reconstructed shower core, and the gamma-hadron separation parameter, respectively}.  All parameter distributions were processed to have mean of zero and standard deviation of unity. This was done both to ensure that the model weights remain small, and to get a useful loss signal from the discriminator early in training (without data standardization, the discriminator would immediately learn the range of the real data and saturate, preventing further training of the generator).

 We designed our model to return a single sample from the 8 dimensional reconstruction distribution8 (the output of the generator is a 1 x 8 vector).  The discriminator was designed to return a loss based on a single sample-to-sample basis, meaning D accepted a batch of N samples in the form of an N x 1 x 8, and returned losses in the form of an N x 1 x 1 tensor.  We chose to do this in order to strongly enforce the joint probability distribution between the elements of the generator sample.
 
For the non-conditional GAN, denoted G, we trained the generator and discriminator with symmetrical models but with different activation functions. Both models have 4 fully-connected layers with 512 nodes in the first layer, 256 in the next two layers. The generator contains 8 nodes in the last layer (one for each parameter) and the discriminator contains one node to represent a probability value. The latent space vector contained 50 elements and was sampled from a normal distribution of zero mean and standard deviation of 1. This model was trained for 10000 total epochs with a learning rate of .03 and a batch size of 2048.

Quantitative validation for GANs can be difficult due to the lack of interpretability of the loss function. We chose to evaluate our 1D GAN results using the  Kullback–Leibler (KL) divergence and the Kolmogorov–Smirnov (KS) test.

The discrete KL divergence is a measure of how different two distributions are from each other on a given, fixed range.  In the field of machine learning it is often regarded as a measure of relative informational entropy of one distribution compared to another.  The KL divergence from distribution $Q$ to distribution $P$ is given by 

\begin{equation}
D_{KL}(P || Q) = - \sum_{i} P(i)\text{log}\left(\frac{Q(i)}{P(i)}\right)
\end{equation}

The KS-test measures the largest bin by bin difference between two normalized CDFs. The closer the difference is to zero, the more likely the two samples come from the same source distribution. 

\begin{figure}[H]
      \includegraphics[width=\textwidth]{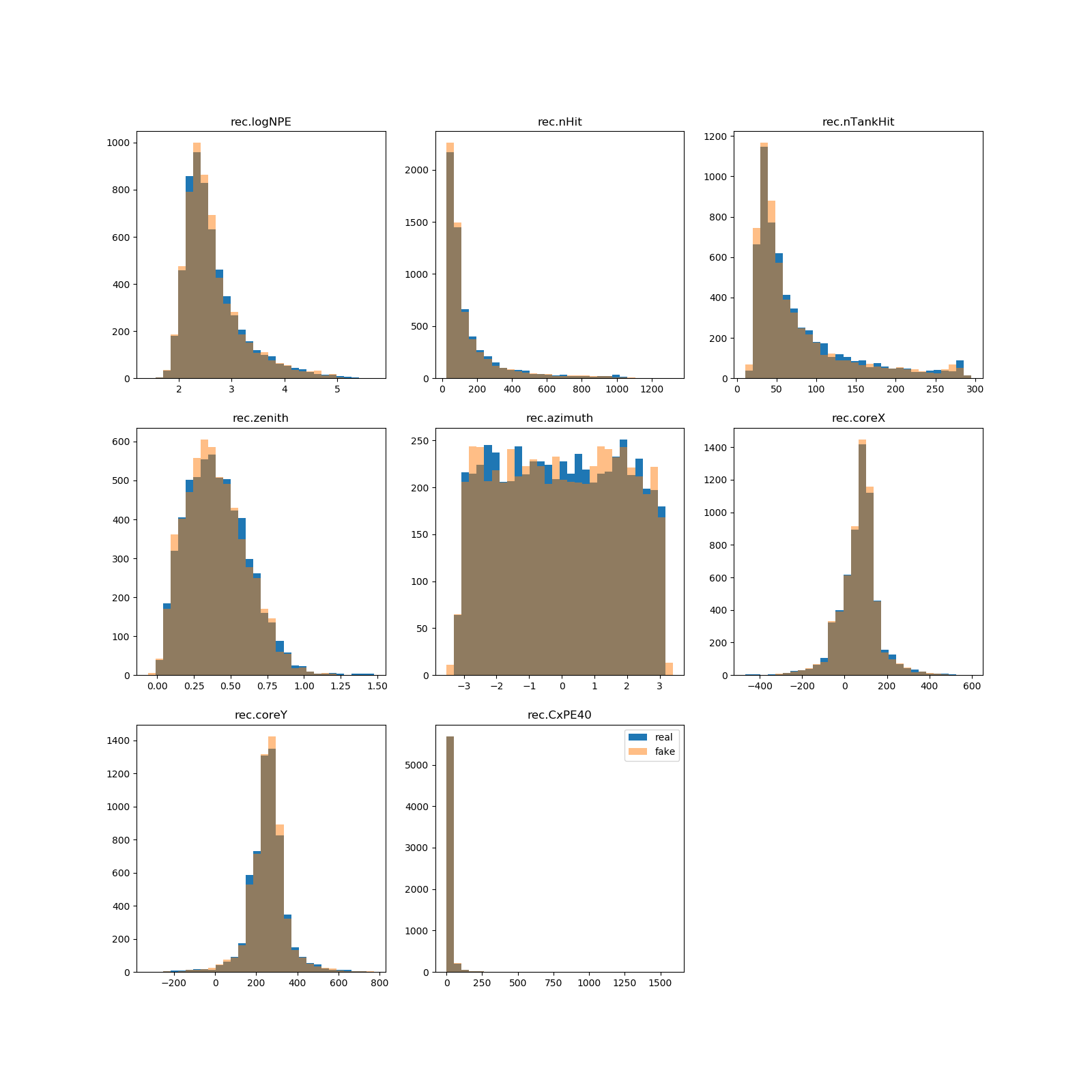}
      \caption{Real (Blue) and Fake (Orange) Distributions. Each histogram represents over 6000 samples from a specific distribution in rec. All real data is unnormalized and represents the reconstruction data in the HAWC simulation pipeline. The orange histogram is slightly transparent to make overlap more apparent.}
      \label{1dhista}
\end{figure}

\begin{figure}[H]
      \centering
      \includegraphics[scale=0.4]{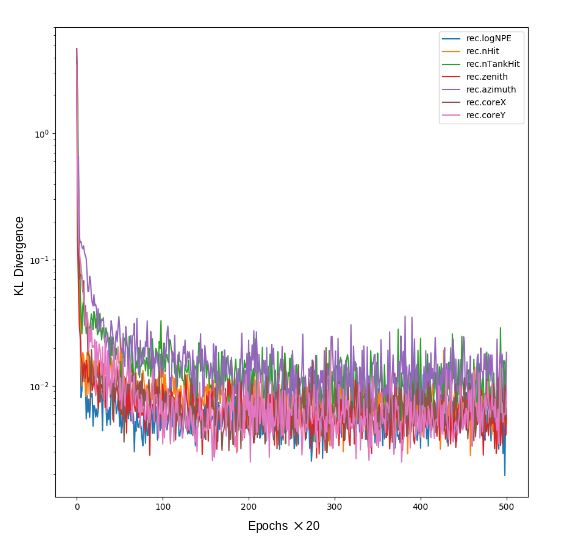}
      \caption{KL divergence over time. Larger values on this plot represent more difference between the generated distribution and the real distribution. }
      \label{1dhistb}
\end{figure}

\subsubsection{Conditional GAN}
The conditional GAN was used to generate new reconstructed simulations with high conditional probability given some input parameters. The conditional parameters were SimEvent.energyTrue, SimEvent.thetaTrue, and SimEvent.phiTrue. These labels were appended to the latent vector of the generator and to the input of the discriminator during training time. The same event was used to get the "real" data rec parameters.

The generator was trained to output samples that would be considered likely given the conditions passed into it. Once the conditional GAN was trained, to generate a sample, a latent vector and a conditional label from each of the SimEvent parameters is passed in.

It is important to note that the conditional variables in the DAQ sim come from some distribution. If the conditional variables are sampled from some other distribution, say uniform on some range, the output distribution will be different. This has an interesting side effect of showing how the GAN is accounting for the conditional inputs: if one of the conditional distributions passed into the generator is the same as a output distribution, the GAN should be acting as an identity function on that variable. We see this in the Azimuth distribution in Figure \ref{fig:1dhistcondb}.
\pagebreak

The conditional GAN was trained on the same hyper parameters as the traditional GAN. Training on a Nvidia GeForce GTX 1080 Ti with 11GB of memory takes ~30 minutes. The generator can generate 15029 [$8\times1$] samples in .002 seconds, but is subject to change depending on network size and latent space size. 
 \begin{figure}{}
      \centering
      \includegraphics[width=\textwidth]{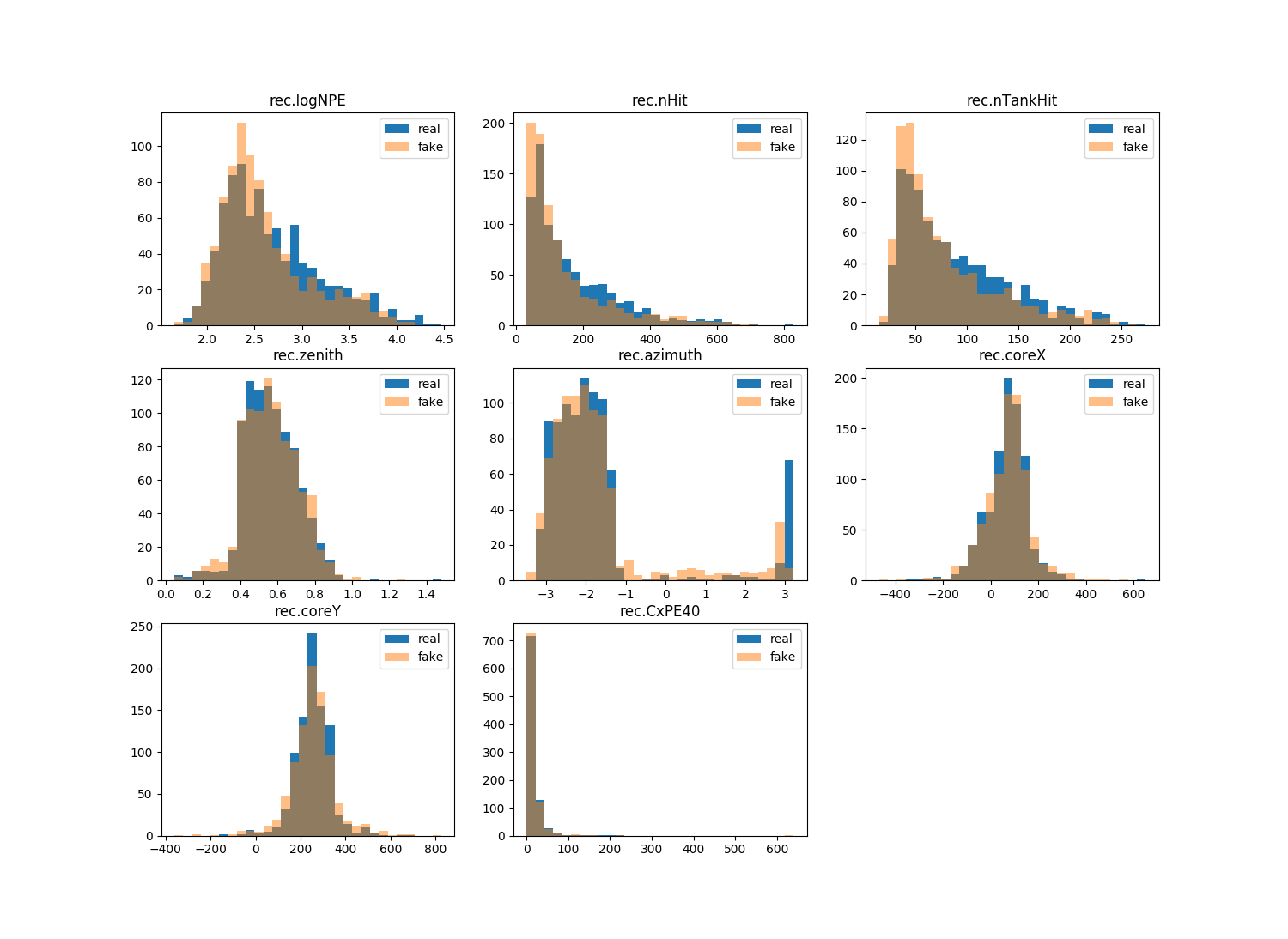}
      \caption{Conditioned data with constrained true conditional distribution. The real distribution (blue) is the rec parameters that correspond to simevent parameters constrained to some range. The generated distribution (orange) represents rec parameters with conditional samples from the simevent distributions.}
      \label{fig:1dhistconda}
\end{figure}
\begin{figure}{}
  \centering
  \includegraphics[width=\textwidth]{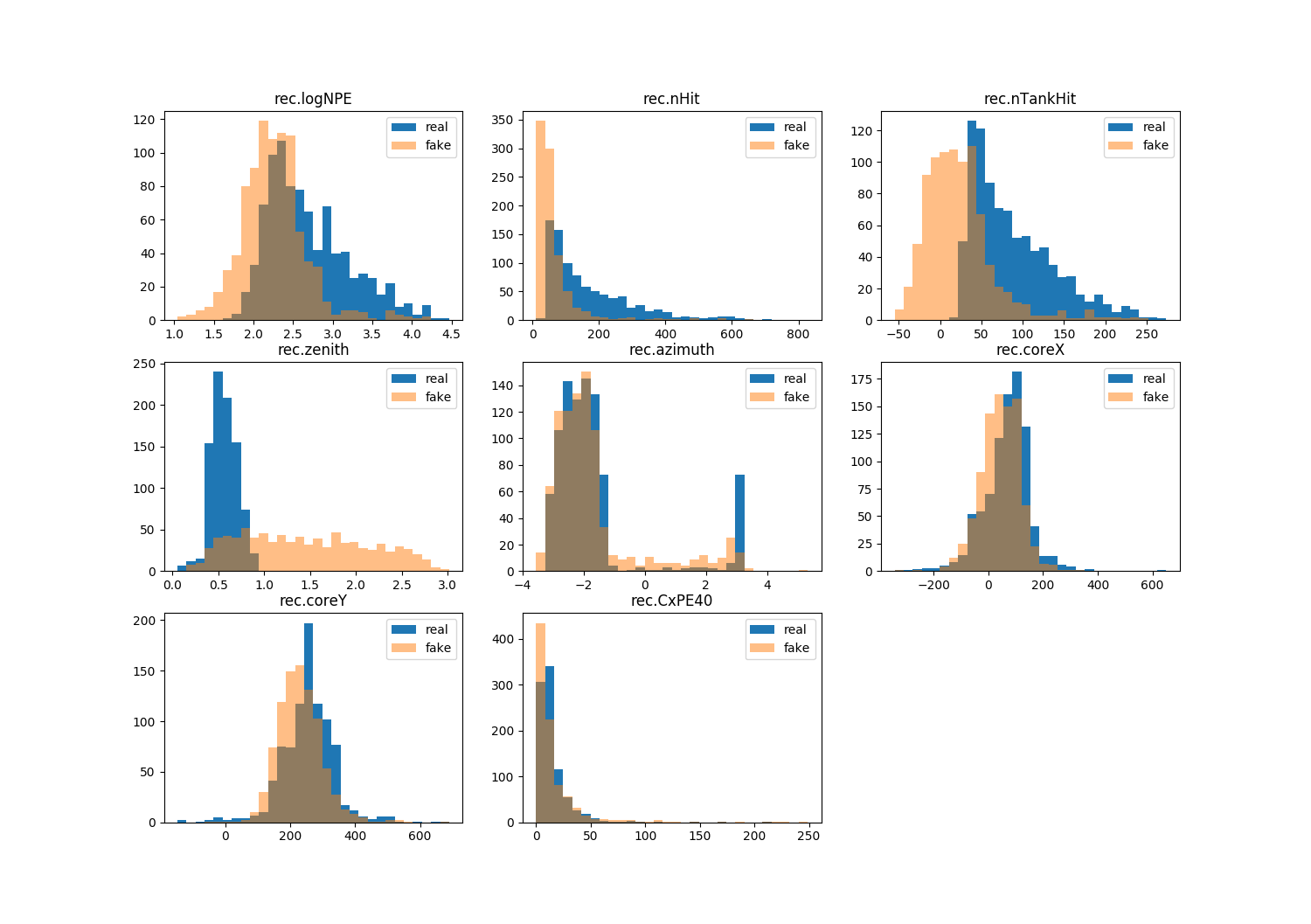}
  \caption{Conditioned data with constrained uniform conditional distribution. The real distribution (orange) is the same as Figure \ref{fig:1dhistconda}. The generated distribution (orange) represents rec parameters with conditional samples from a uniform distribution.}
  \label{fig:1dhistcondb}
\end{figure}

\subsection{2D Data Generation with PixelCNN}
\subsubsection{PixelCNN Model Modifications}
We trained our PixelCNN model to minimize discrete log likelihood loss, which is proportional to bits per dimension. Bits per dimension can be interpreted as the number of bits needed to compress every subpixel value using this model as a compression scheme. With a lower number, we can think of the model as being able to represent more of the features of images in the distribution we are trying to learn. These will be encoded by the parameters in the model itself, and are not reflected in the final size of the compressed image. We reported results in bits per dimension, so they can be compared to those of other similar models. 

We modified the subpixel computation to adjust the model for the HAWC dataset. Recall that for each PMT, we need to sample both time and intensity values.

\begin{equation}
p(x_i) = p(x_{i,\text{intensity}}|x_{<i})p(x_{i,\text{time}}|x_{<i,x_{i,\text{intensity}}}))
\end{equation}

We additionally modified this equation in a similar fashion to \cite{PixelCNNpp}, where the value for the time channel is a mixture of logistics that linearly depend on the intensity channel. This allowed us to sample additional channels per pixels without increasing the complexity or runtime significantly. While the dependency between time and intensity is not as simple as that between RGB values, empirically we found that our model was expressive enough to learn good distributions for both channels.

\subsubsection{Training Procedure}
We followed the standard approach used in \cite{PixelCNNpp}, and did not modify any of the hyperparameters of their model (\url{https://github.com/openai/pixel-cnn}), so faster convergence might be possible with some tuning.

We used a neural network architecture consisting of stacked ResNet \cite{ResNet} blocks, defined in \cite{PixelCNN}. The convolutions were casual, with a boolean mask applied to ensure computation on one pixel is only dependent on pixels that were already previously generated. For more details on the architecture, see \cite{PixelCNN}.

We trained the conditional, two-channel model on a single Tesla V100 in about three and a half hours or five epochs. We were able to train to absolute convergence within 50 epochs, for a marginal improvement in the metric, but with approximately equal visual quality. All the other models took roughly the same amount of time to train, on the same GPU. See Table \ref{tab:pixelcnn-params} for the hyperparameters used in training the model.

\begin{table}[htb]
  \centering
  \begin{tabular}{ll}
    \toprule
    \cmidrule(r){1-2}
    Parameter & Value \\
    \midrule
    Learning Rate & 0.001 \\
    Learning Rate Decay & 0.999995 \\
    Batch Size & 16 \\
    Dropout Probability & 50\% \\
    Number of filters & 160 \\
    Logistic components & 10 \\
    Residual blocks per stage & 5 \\
    Seed & 1 \\
    \bottomrule
  \end{tabular}
  \caption{Hyperparameters used for all PixelCNN results, with model from \cite{PixelCNNpp}}
  \label{tab:pixelcnn-params}
\end{table}

\subsubsection{Conditional PixelCNN} \label{sec:cond-pixelcnn}
We were able to create a conditional variant of the PixelCNN model using the procedure described in \cite{condPixelCNN}, where the variable(s) we want to condition on were added to the activation functions in the network. When the conditional model was trained on the ImageNet \cite{imagenet} and CIFAR10 \cite{CIFAR10} datasets, it was able to generate images belonging to a specific class by being fed a vector corresponding to the class. 

We can think of this as a modification to equation \ref{eq:pixelcnn}, where the terms are dependent on a latent vector $\mathbf{h}$:

\begin{equation}
p(\mathbf{x} | \mathbf{h}) = \prod_{i=1}^{n^2}p(x_i | x_1, \dots , x_{i_1}, \mathbf{h})
\end{equation}

\begin{figure}[H]
  \centering
  \includegraphics[scale=0.3]{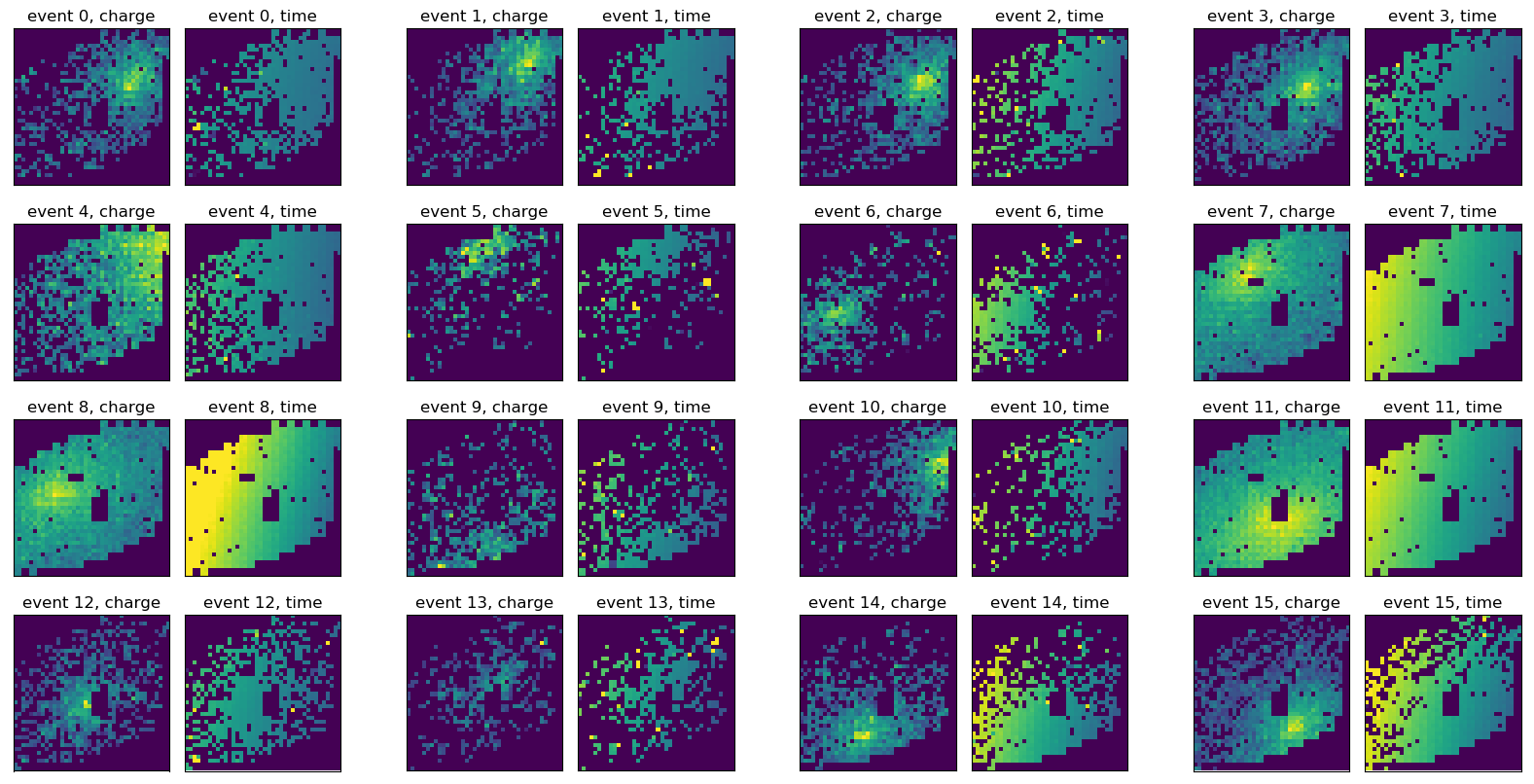}
  \caption{Conditional, two channel data with azimuth=0.0, cherry-picked from samples with larger hits for clarity. Note that if picked randomly, the samples would have the sample distribution of sparse and dense events as displayed in the other plots in the Appendix. We can see a clear affect of fixing azimuth in the time channel, where the angle at which shower moves across the grid is clearly represented in the gradient.}
  \label{fig:azimuth_0}
\end{figure}

\begin{figure}[H]
  \centering
  \includegraphics[scale=0.3]{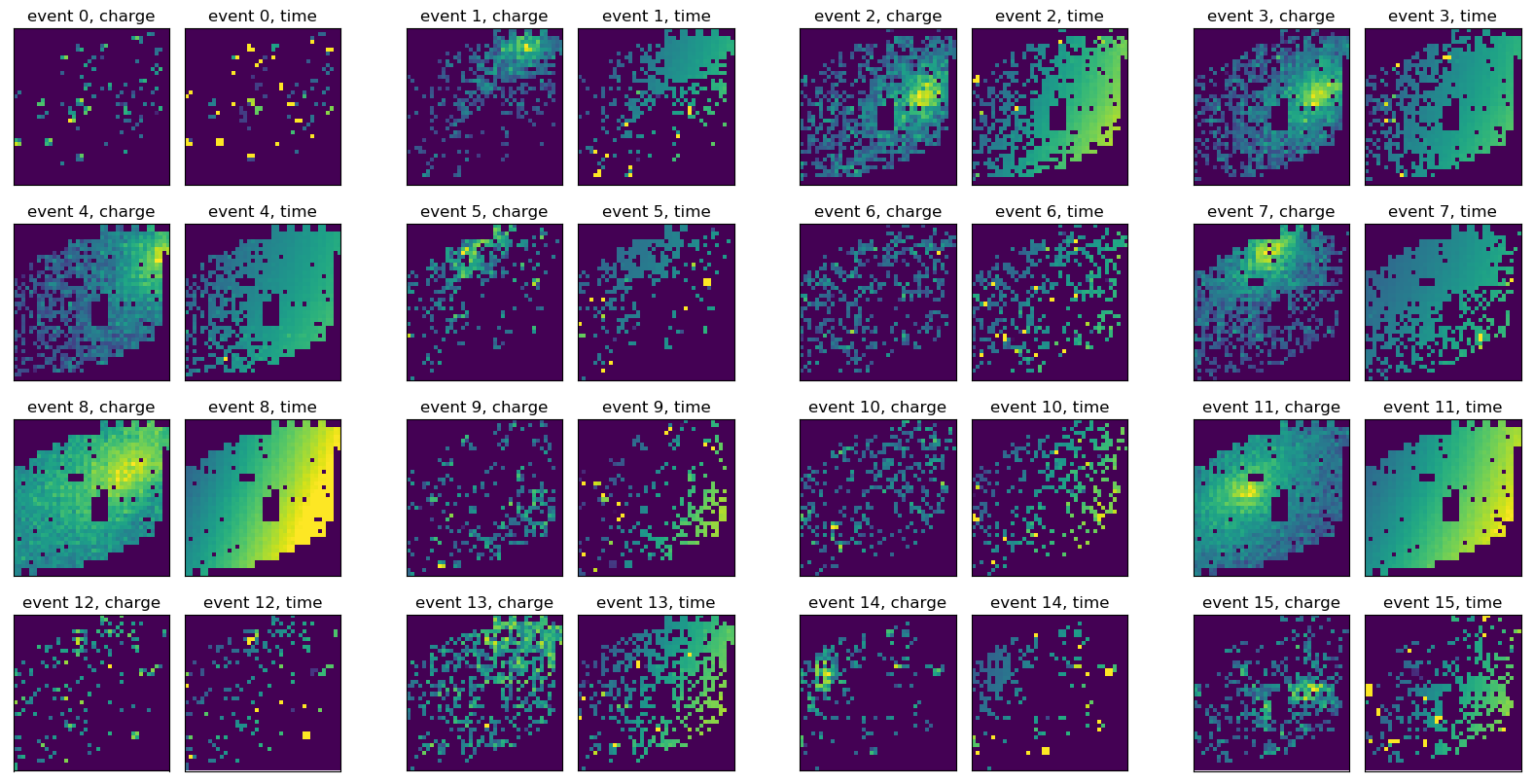}
  \caption{Conditional, two channel data with azimuth=3.0, cherry-picked from samples with larger hits for clarity. We can see a clear affect of fixing azimuth in the time channel, where the angle at which shower moves across the grid is clearly represented in the gradient. Note the contrast in angles of the gradient between this plot and Figure \ref{fig:azimuth_0}.}
  \label{fig:azimuth_3}
\end{figure}

We performed an experiment where a conditional PixelCNN model was trained on the HAWC dataset, conditioned on the value of Azimuth (rec.Azimuth). This allows us to generate data only from a specific value of Azimuth. Generated images should appear distributed similar to the distribution of data with the same Azimuth value found in the dataset. See Figures \ref{fig:azimuth_0} and \ref{fig:azimuth_3} for an example of this conditioning. Note that we are able to condition on an arbitrary number of variables, so by conditioning on all the variables used in the HAWC simulation, it should be possible to generate images from any set of desired input parameters.

\subsubsection{Fast Sampling} \label{sec:fast-sampling}

Using improvements from \cite{fastPixelCNN}, we were able to achieve orders of magnitude of improvement in the speed of image generation. Although we only had access to a 1080 Ti workstation while testing the fast model, from our tests we found that the time per batch of samples did not increase at all compared while increasing the batch size. Comparatively, with the original model we saw a linear increase of sample time relative to batch size.

The maximum batch size we could fit in the memory of a 1080 Ti was 512, but we also showed projected results for other systems with more memory. Note that our projected results only accounted for the increase in memory, and not the increase in GPU clock speed or performance relative to our 1080 Ti. We show results in Table~\ref{tab:fast-pixelcnn}.

\begin{table}[htb]
  \centering
  \begin{tabular}{llll}
    \toprule
    \cmidrule(r){1-2}
    Model     & GPU     & Time/Event(s) & Samples/s \\
    \midrule
    PixelCNN [ref] & Tesla V100 16 GB  & 17    & 0.059 \\
    Fast PixelCNN [ref]     & GTX 1080 Ti 11GB & 0.17    & 5.8 \\
    Fast PixelCNN (projected)     & Tesla V100 32 GB       & \textless 0.05  & 20  \\
    Fast PixelCNN (projected)    & DGX-2 (512 GB)       & \textless 0.0017  & 580 \\
    \bottomrule
  \end{tabular}
  \caption{Event generation speed comparison between GPU and model changes}
  \label{tab:fast-pixelcnn}
\end{table}

It would be possible to sample at an even faster rate using TensorRT to simplify the compute graph or by using low, 8-bit precision. This would be reasonable considering the ground truth data only holds 1 decimal place of precision for all results.

\begin{figure}[H]
  \centering
  \includegraphics[scale=0.3]{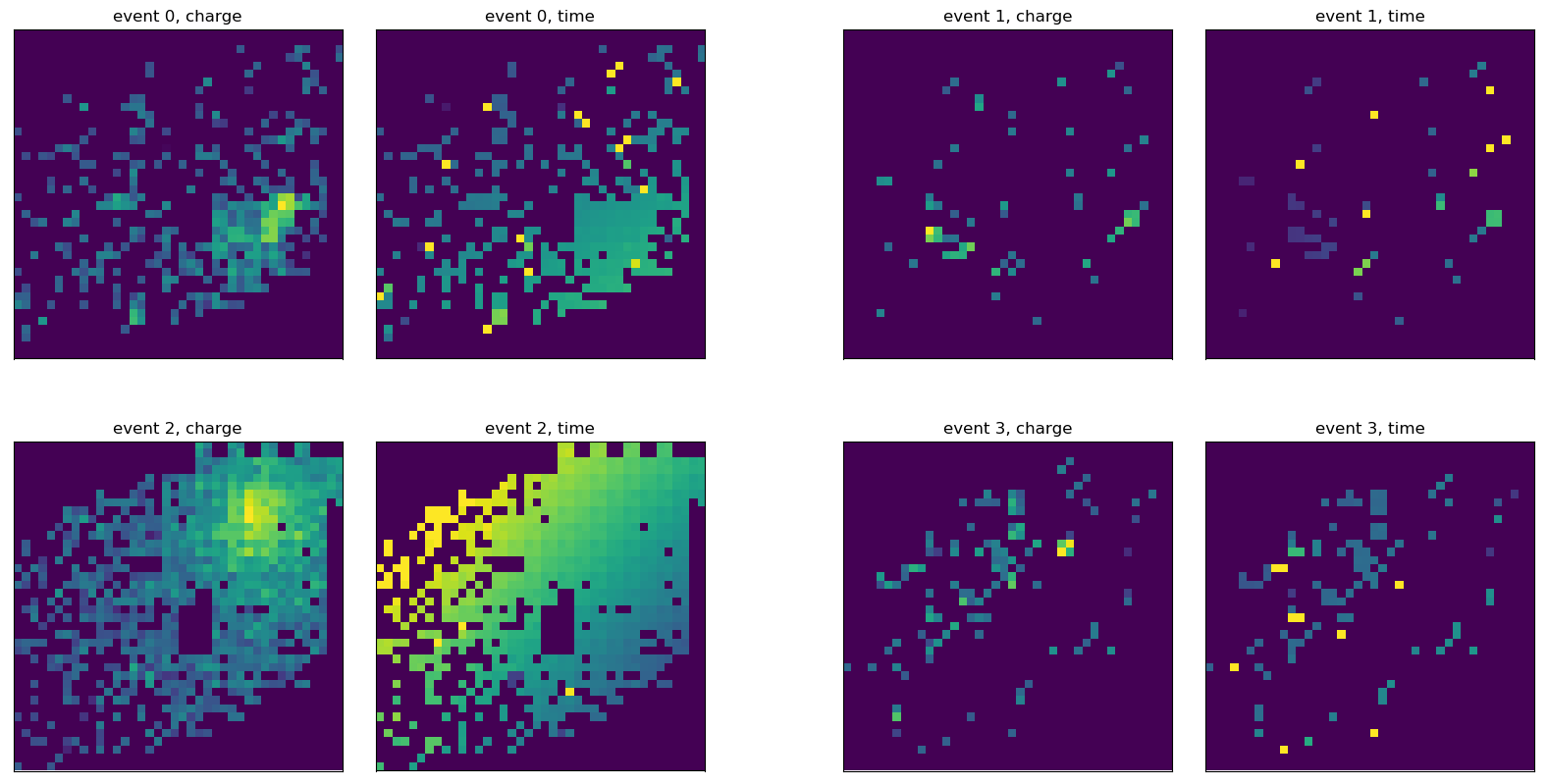}
  \caption{Samples from fast, unconditional two-channel PixelCNN.}
  \label{fig:fast-pixelcnn-sample}
\end{figure}

Importantly, the quality of the samples did not seem to decrease from visual inspection (see Figure~\ref{fig:fast-pixelcnn-sample}).

\subsection{2D Data Generation with GANs}
\subsubsection{2D GAN}

Our 2D GAN model functions in the the same way as the 1D GAN except the data passed in is images. The change in architecture is simple; the linear layers were replaced with convolutional and deconvolutional layers in the  discriminator and generator, respectively.

The $40\times40$ images input into this pipeline have two channels (charge and time) representing PMT data of a specific event. Because the images have 1600 pixels and there are only 1200 PMT in the HAWC Observatory, many pixels on the side and in the middle of the training images are always zero.

For this dataset, the generator consisted of a linear layer followed by 2 nearest neighbor upsampling and convolution layers and ended with 3 convolution layers. In addition, to help with training stability, the Wasserstein-1 loss function was used and the sigmoid activation function at the end of the discriminator was removed. 

\begin{table}[htb]
  \centering
  \begin{tabular}{ll}
    \toprule
    \cmidrule(r){1-2}
    Parameter & Value \\
    \midrule
    Learning Rate & 1e-4 \\
    Batch Size & 64 \\
    Latent Size & 128 \\
    Number of Layers & 5 \\
    Symmetric GAN? & True\\
    Gradient Penalty ($\lambda$) & 1 \\
    \bottomrule
  \end{tabular}
  \caption{Hyperparameters used for 2D Wasserstein GAN.}
  \label{tab:2DGAN-params}
\end{table}

Training on a Nvidia GeForce GTX 1080 Ti with 11GB of memory takes 2+ hours for 10,000 epochs. Inference time takes .03 to .05 seconds for 2000 images, depending on number of channels, size of the entropy source, and size of the generator. 


\section{Results and Conclusions}
\subsection{1D Generation}

\subsubsection{1D GAN}

The results, Figure \ref{1dhista}, show that the generated data closely resembles the original distributions at a qualitative level. 

Figure \ref{1dhistb} shows the KL divergence over time. The interpretation of this plot is that the GAN is able to make the generated distributions closer to the real distributions over time.

Since the dimensionality of the output is small compared to other types of data generated by GANs (such as images or text), the 1D GAN had no problems with stability or in learning the 8 unique distributions. Training time took approximately 30 minutes, but it is recommended to train for 1-2 hours to ensure convergence.

With 60,000 samples each from real and fake distributions, we computed the 2-sample KS-statistic:

\begin{table}[htb]
  \centering
  \begin{tabular}{lllll}
    \toprule
    \cmidrule(r){1-2}
    Distribution & KS-Statistic \\
    \midrule
    rec.logNPE & 0.0316\\
    rec.nHit & 0.0129\\
    rec.nTankHit & 0.0488\\
    rec.zenith & 0.0110\\
    rec.azimuth & 0.0110\\
    rec.coreX & 0.0129\\
    rec.coreY & 0.0132\\
    rec.CxPE40 & 0.0072\\
    \bottomrule
  \end{tabular}
  \label{tab:kstest-results}
\end{table}

\subsection{2D Generation}

\subsection{PixelCNN}

See Table~\ref{tab:pixelcnn-results} for visual results. As a reference, the same PixelCNN model trained on the CIFAR 10 dataset achieved a bits per dimension of 2.92. The results we achieved, of about 0.71 bits per dimension, seemed reasonable since our data is not as complex as data from electro-optical sensors, and not as diverse in distribution. We could expect variation in sparsity, angle of the shower, etc., but not differences in classes like cats and airplanes.

\begin{table}[htb]
  \centering
  \begin{tabular}{lllll}
    \toprule
    \cmidrule(r){1-2}
    Dataset & Conditional     &  Epochs     & Time(hr)  & Bits/dim \\
    \midrule
    CIFAR 10 \cite{PixelCNNpp} & No & - & - & 2.92 \\
    CIFAR 10 \cite{PixelCNNpp} & Yes & - &  - & 2.94 \\
    HAWC & Yes & 5  & 3.0    & 0.7234 \\
    HAWC &Yes     & 38 & 23.0    & 0.7117 \\
    HAWC & No     & 5     & 3.0  & 0.7269  \\
    HAWC & No    & 40  & 24.2  & 0.7135 \\
    \bottomrule
  \end{tabular}
  \caption{Results of PixelCNN models trained on a single Tesla V100 16GB GPU. CIFAR 10 results from original papers. After 5 epochs, we found that bits per dim was already very low, and samples generated by the model looked very realistic. For bits per dim, lower is better.}
  \label{tab:pixelcnn-results}
\end{table}

\begin{figure}[H]
  \centering
  \includegraphics[scale=0.3]{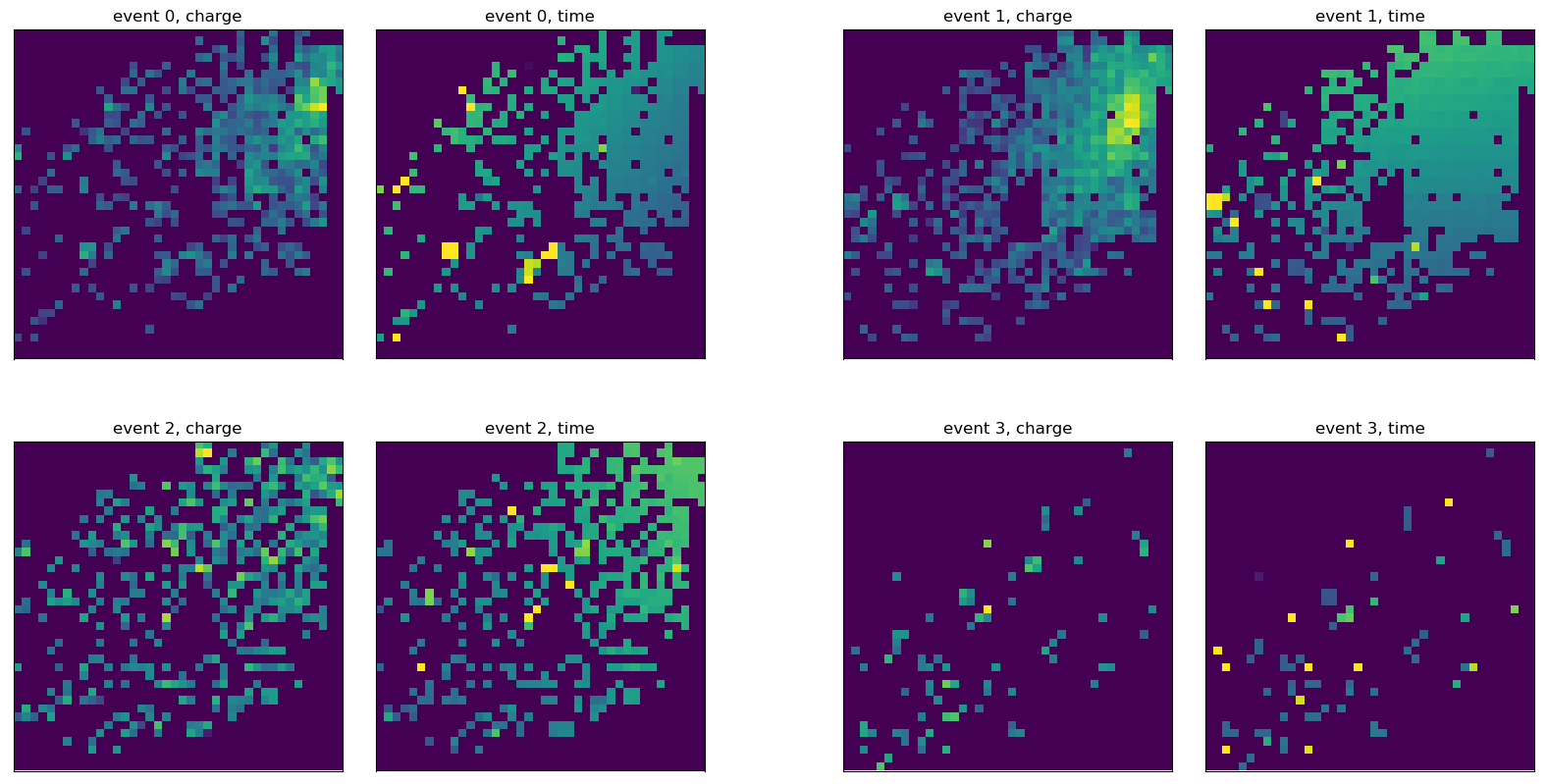}
  \caption{Two Channel data from PixelCNN. We could observe all of the features of the ground truth dataset in these samples. The samples looked very representative of the dataset; the PixelCNN model learned with perfect accuracy that ``dead'' pixels on the edges and center stay unlit, and PMT hits in one channel also result in hits in the other channel. For additional visuals, see the Appendix.}
\end{figure}

\subsubsection{2D GAN}
\begin{figure}[H]
  \centering
  \includegraphics[width=\textwidth]{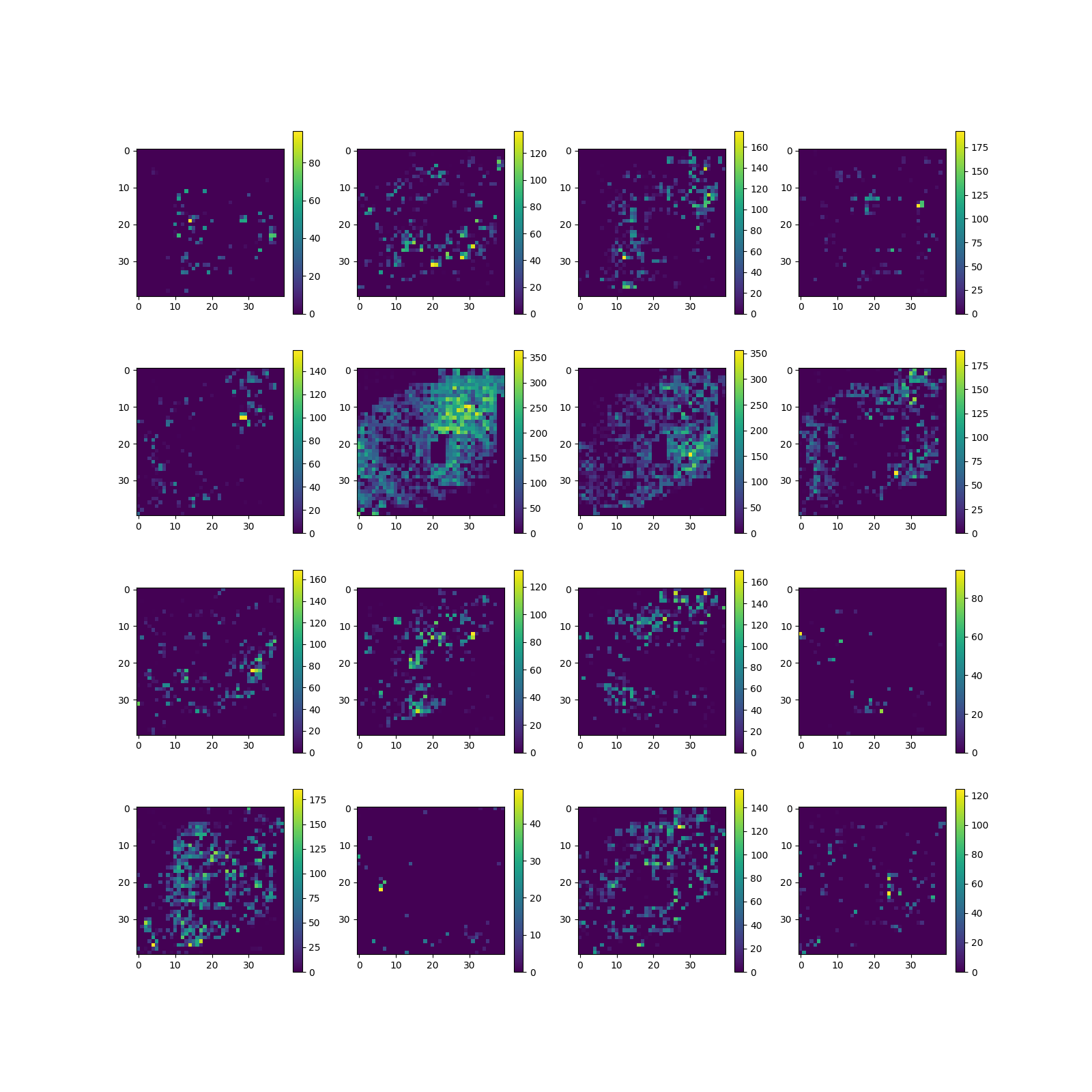}
  \caption{1-Channel generated images of charge from 2D GAN. }
  \label{fig:2dgan1channel}
\end{figure}

The WGAN was able to adequately learn the image distributions of the charge channel, but struggled to learn both charge and time channels at the same time. While the same pixels in both channels were on, the generator never learned to make a smooth gradient in the time channel. One channel images of charge are shown in Figure \ref{fig:2dgan1channel}.

The WGAN proved to have learned the areas that are always zero pixel value (denoted by no PMT in that pixel), particularly in the top left, bottom right, and center areas. In addition, the Wasserstein distance as expected decreases over time, showing that the generator is learning to match the image distributions. 

\section{Acknowledgements}
We the authors would like to thank the HAWC collaboration for generously providing us with the simulations we needed for this work.  We would specifically like to thank Edna Loredana Ruiz Velasco of the HAWC collaboration, who provided the initial pipeline for processing the HAWC XCDF files.

The codebase for this work can be found at \url{https://github.com/arcelien/hawc-deep-learning}

\bibliographystyle{plain}
\bibliography{references}

\section{Appendix}

\subsection{Additional 2D Visualizations}
We show additional samples from various PixelCNN models.

\begin{figure}[H]
  \centering
  \includegraphics[scale=0.3]{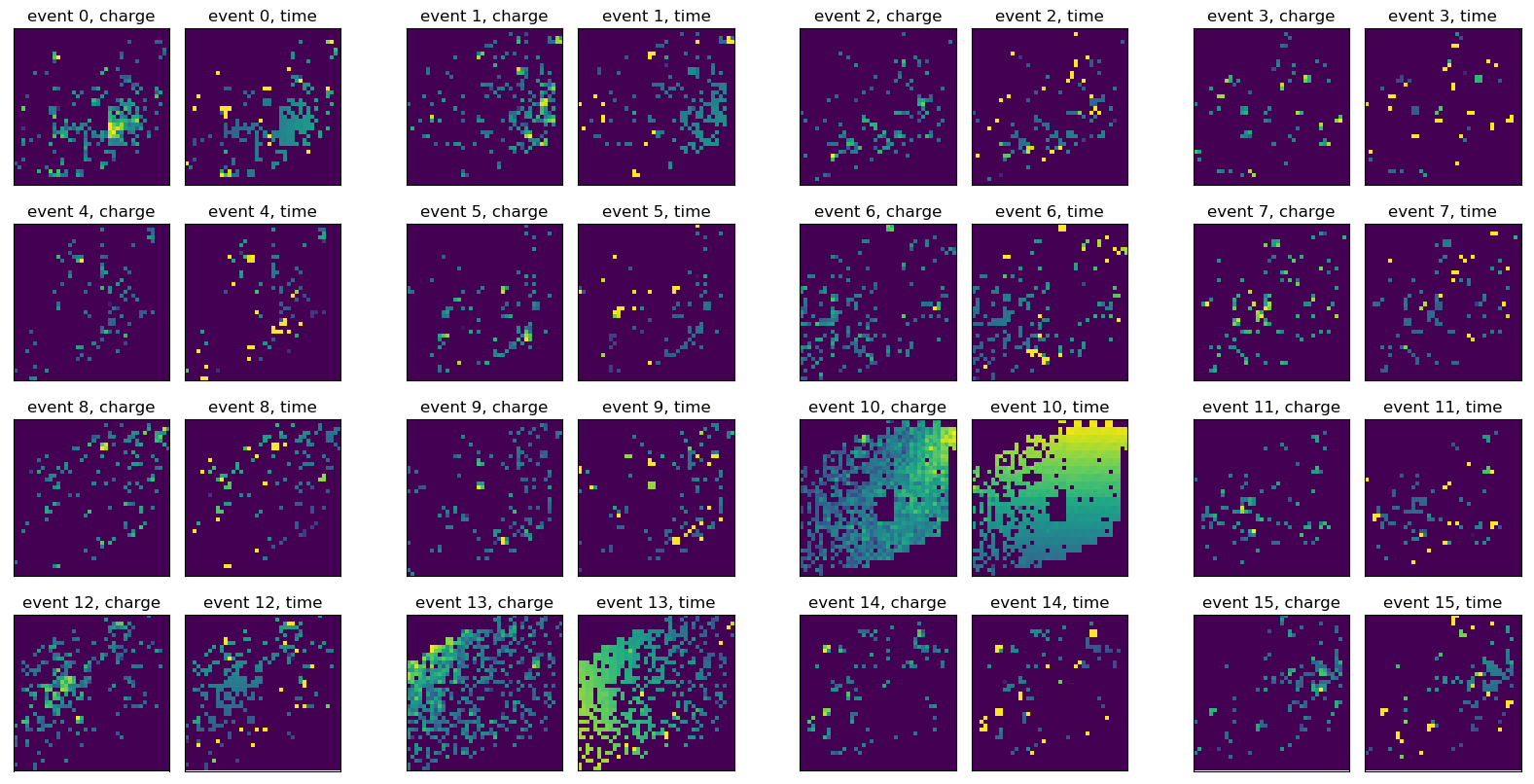}
  \caption{Additional visualizations from the unconditional PixelCNN model.}
\end{figure}

\begin{figure}[H]
  \centering
  \includegraphics[scale=0.3]{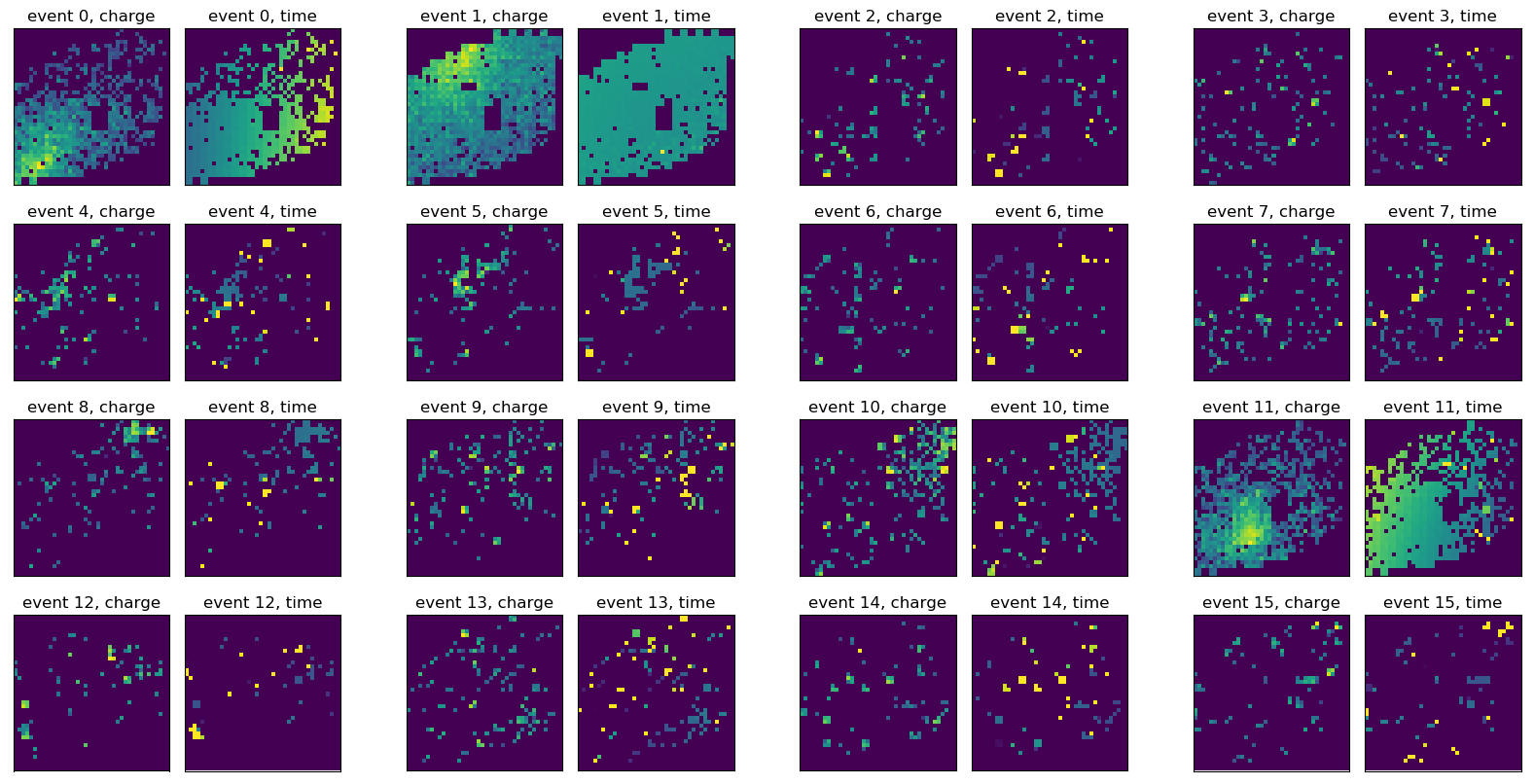}
  \caption{Additional visualizations from the fast, unconditional PixelCNN model}
\end{figure}

\begin{figure}[H]
  \centering
  \includegraphics[scale=0.4]{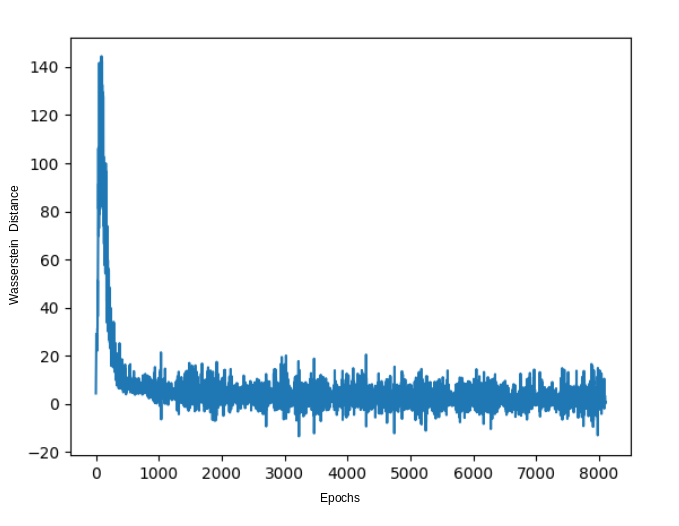}
  \caption{Estimated Wasserstein Distance over time. The discriminator was initially easily able to discriminate samples because it was trained for more iterations than the generator. The generator gradually improved over time and resulted in an interpretable loss curve that decreases in the long term.}
\end{figure}

\end{document}